\documentclass[aps,amssymb]{revtex4}                  
\usepackage{graphicx}
\usepackage{amsmath}
\newcommand{\tr}{CMR }
\newcommand{\dfk}{TRFK }
\newcommand{\s}{blue }
\newcommand{\w}{red }

\newcommand{\T}{{\cal T }}

\begin{document}
\title{A Percolation-Theoretic Approach to Spin Glass Phase Transitions

}

\author{J.~Machta}
\affiliation{
Physics Department,
University of Massachusetts,
Amherst, MA 010003 USA}
\email{machta@physics.umass.edu}
\author{C.M.~Newman}
\affiliation{
Courant Institute of Mathematical Sciences,
New York University,
New York, NY 100012 USA}
\email{newman@courant.nyu.edu}     
\author{D.L.~Stein}
\affiliation{
Physics Department and Courant Institute of Mathematical Sciences,
New York University,
New York, NY 100012 USA}
\email{daniel.stein@nyu.edu}
\begin{abstract}
The magnetically ordered, low temperature phase of Ising ferro-
magnets is manifested within the associated Fortuin-Kasteleyn (FK) random cluster representation by the occurrence of a single positive density
percolating cluster. In this paper, we review our recent work on the
percolation signature for Ising spin glass ordering --- both in the
short-range Edwards-Anderson (EA) and  infinite-range
Sherrington-Kirkpatrick (SK) models --- within a two-replica FK
representation
and also in the different Chayes-Machta-Redner two-replica graphical
representation. Numerical studies of the $\pm J$ EA model in dimension
three and
rigorous results for the SK model are consistent in supporting the
conclusion that the signature of spin-glass order in these models is the
existence of a single percolating cluster of maximal density normally
coexisting with a second percolating cluster of lower density.
\end{abstract}
\maketitle

\section{Introduction}
\label{sec:intro}

The question of whether laboratory spin glasses --- or the theoretical
models used to represent them --- have a thermodynamic phase transition
remains unresolved despite decades of work~\cite{BY86,MPRRZ00,NSjpc03}.
Although the infinite-range Sherrington-Kirkpatrick~(SK) Ising spin glass
is easily shown to possess a phase transition~\cite{SK75,ALR87}, the
existence of one in the corresponding short-range
Edwards-Anderson~(EA)~\cite{EA75} Ising model (on the cubic lattice ${\bf
Z}^d$) has not been established in {\it any\/} finite dimension.  While
some evidence for a transition has been uncovered through high-temperature
expansions~\cite{FS90,TH96}, analytical studies of variable long-range $1d$
models~\cite{KAS83}, and extensive numerical
simulations~\cite{BY86,O85,OM85,KY96}, the issue remains unresolved~\cite{MPR94}.

Random graph methods, and in particular the Fortuin-Kastelyn~(FK) random
cluster~(RC) representation~\cite{KF69,FK72}, provide a set of useful tools
for studying phase transitions (more specifically, the presence of multiple
Gibbs states arising from broken spin rotational symmetry) in discrete spin
models.  In these representations spin correlation functions can be
expressed through the geometrical properties of associated random graphs.
FK and related models are probably best known in the physics literature for
providing the basis for powerful Monte Carlo methods for studying phase
transitions~\cite{Swe,SwWa86,Wolff}, but they have also proved important in
obtaining rigorous results on phase transitions in discrete-spin
ferromagnetic (including inhomogeneous and randomly diluted)
models~(e.g.,~\cite{ACCN87,IN88}). Because of complications due to
frustration, however, graphical representations have so far played a less
important role in the study of spin glasses.

The goal of the studies presented here is to construct a viable approach
which uses random graph methods to address the problem of phase transitions
and broken symmetry in spin glass models.  In earlier
papers~\cite{MNS07a,MNS07b}, we studied the ``percolation signature'' of
spin glass ordering within two different graphical representations --- the
two-replica model of Chayes, Machta and Redner~(CMR)~\cite{CMR98,CRM98} and
a two-replica version of the FK~representation~(TRFK), as proposed in
Sec.~4.1 of~\cite{NS07}.  The purpose of that analysis was to show that FK
methods could be utilized to study spin glass phase transitions.  The
result of this work was the uncovering of strong evidence that the
existence of a spin glass transition coincides with the emergence of {\it
doubly\/} percolating clusters of {\it unequal\/} densities.  This scenario
is more complex than what occurs in ferromagnetic models, where the phase
transition coincides with percolation of a single FK cluster.

In what follows, we first review the FK random cluster representation for
ferromagnetic models, and then describe our analytical and numerical
results for spin glasses.

\section{The Fortuin-Kasteleyn Random Cluster Representation}
\label{sec:FK}

In this section we briefly review the Fortuin-Kasteleyn random cluster
representation~\cite{KF69,FK72} which relates the statistical mechanics of
Ising (or Potts) models to a dependent percolation problem.  Our focus will
be on Ising models throughout, but the analysis is easily extended to more
general Potts models.

\subsection{Ferromagnetic Models}
\label{subsec:ferro}

We start by considering a nearest-neighbor Ising ferromagnet, whose
couplings $J_{xy}\ge 0$ are not necesarily identical:

\begin{equation}
\label{eq:EA}
{\cal H}=-\sum_{\langle xy\rangle} J_{xy}\sigma_x\sigma_y\ ,
\end{equation}
where, as already noted, $J_{xy}\ge 0$ and the sum is over nearest neighbor
pairs of sites in ${\bf Z}^d$.  Each such coupling is associated with an
edge, or bond, $\langle x,y\rangle$, with the set of all such bonds denoted
${\bf E}^d$.

The RC approach introduces parameters $p_{xy}\in[0,1)$ through the formula:
\begin{equation}
\label{eq:pxy}
{\cal P}_{xy}=1-\exp[-2\beta|J_{xy}|]\, ,
\end{equation}
where $\beta$ is the inverse temperature.  One can then define a
probability measure $\mu_{\rm RC}$ on $\{0,1\}^{{\bf E}^d}$, that is, on 0-
or 1-valued bond occupation variables $\omega_{xy}$.  It is one of two
marginal distributions (the other being the ordinary Gibbs distribution) of
a joint distribution on $\Omega=\{-1,+1\}^{{\bf Z}^d}\times\{0,1\}^{{\bf
E}^d}$ of the spins and bonds together (such a joint distribution will be
introduced in Sec.~\ref{subsec:CMR}).  The marginal distribution $\mu_{\rm
RC}$ is given formally by
\begin{equation}
\label{eq:mufk}
\mu_{\rm RC}(\{\omega_{xy}\})=Z_{\rm RC}^{-1}\ 2^{\#(\{\omega_{xy}\})}\
  \mu_{\rm ind}(\{\omega_{xy}\})\ 1_U(\{\omega_{xy}\})\, ,
\end{equation}
where $Z_{\rm RC}$ is a normalization constant, $\#(\{\omega_{xy}\})$ is
the number of clusters determined by the realization $\{\omega_{xy}\}$,
$\mu_{\rm ind}(\{\omega_{xy}\})$ is the Bernoulli product measure
corresponding to independent occupation variables with $\mu_{\rm
ind}(\{\omega_{xy}=1\})={\cal P}_{xy}$, and $1_U$ is the indicator function
on the event $U$ in $\{0,1\}^{{\bf E}^d}$ that there exists a choice of the
spins $\{\sigma_x\}$ so that $J_{xy}\omega_{xy}\sigma_x\sigma_y\ge 0$ for
all $\langle x,y\rangle$~\cite{WaSw88,KO88,N93}.

There are several things to be noted about~(\ref{eq:mufk}).  The factor 2
in the term $2^{\#(\{\omega_{xy}\})}$ arises because we've confined
ourselves to Ising models, so that every connected cluster of spins (each
such connected cluster consists of all satisified bonds) can be in one of
two states (in the ferromagnetic case, all up or all down); in a $q$-state
Potts model, this term would then be replaced by $q^{\#(\{\omega_{xy}\})}$.
More importantly, the indicator function on $U$, which is the event that
there is no frustration in the occupied bond configuration, is always one
for the ferromagnet; consequently, this term is superfluous for
ferromagnetic models.  We include it, however, because it will be needed
when we generalize to models with frustration.  Finally, we note that
finite-volume versions of the above formulas, with specified boundary
conditions, can be similarly constructed.

In the case of a ferromagnet, there are general theorems~\cite{BK89} which
ensure that when percolation occurs, there is a unique percolating cluster.
It then easily follows that RC percolation within the FK representation (or
``FK percolation'' for short) corresponds to the presence of multiple Gibbs
states (in the ferromagnet, magnetization up and magnetization down), and
moreover that the onset of percolation occurs at the ferromagnetic critical
temperature.  To prove this, it is sufficient to show that FK percolation
is both necessary and sufficient for the breaking of global spin flip
symmetry in the ferromagnet. To see that FK percolation is a necessary
condition, note that the contribution to the expectation of $\sigma_0$ from
any finite RC cluster is zero: if a spin configuration $\sigma$ is
consistent with a given RC bond realization within such a cluster, so is
$-\sigma$, and both will be equally likely.  As a consequence,
$\langle\sigma_0\rangle=0$ in infinite volume in the absence of RC
percolation.

To see that RC percolation is a sufficient condition for the magnetization
order parameter to be nonzero, consider a finite volume $\Lambda_L$ with
fixed boundary conditions, i.e., a specification $\overline{\sigma}_x=\pm
1$ for each $\overline{\sigma}_x\in\partial\Lambda_L$.  For the
ferromagnet, by first choosing all $\overline{\sigma}_x=+1$ and then all
$\overline{\sigma}_x=-1$, one can change the sign of the spin $\sigma_0$ at
the origin even as $L\to\infty$.  That is, boundary conditions infinitely
far away affect $\sigma_0$, which is a signature of the existence of
multiple Gibbs states.

When attention is confined to ferromagnetic models, the mapping of the FK
formalism to interesting statistical mechanical quantities is
straightforward (and intuitive); for example 
\begin{equation}
\label{eq:ferropath}
\langle\sigma_x\sigma_y\rangle=\mu_{\rm RC}(x\leftrightarrow y)\, ,
\end{equation}
where $\langle\sigma_x\sigma_y\rangle$ is the usual Gibbs two-point
correlation function and $\mu_{\rm RC}(x\leftrightarrow y)$ is the RC
probability that $x$ and $y$ are in the same cluster.  Similarly, with
``wired'' boundary conditions (i.e., each boundary spin is connected to its
nieghbors), one has
\begin{equation}
\label{eq:ferroperc}
\langle\sigma_x\rangle_+=\mu_{\rm RC}(x\leftrightarrow\infty)\, .
\end{equation}
So for ferromagnets, a phase transition from a unique (paramagnetic) phase
at low $\beta$ to multiple infinite-volume Gibbs states at large $\beta$ is
equivalent to a percolation phase transition for the corresponding RC
measure.

\subsection{Spin Glass Models; The TRFK Representation}
\label{subsec:sgs}

For spin glasses (or other nonferromagnets with frustration) the situation
is more complicated.  Now for two sites $x$ and $y$, (\ref{eq:ferropath})
becomes
\begin{equation}
\label{eq:sgpath}
\langle\sigma_x\sigma_y\rangle=\langle 1_{x\leftrightarrow y}\eta(x,y)\rangle_{RC};
\,\,\,\eta(x,y)=\prod_{\langle x'y'\rangle\in{\cal
C}}{\rm sgn}(J_{x'y'})\, ,
\end{equation}
where ${\cal C}$ is any path of occupied bonds from $x$ to $y$.  By the
definition of $U$, any two such paths ${\cal C}$ and ${\cal C'}$ in the
{\it same\/} cluster will satisfy $\prod_{\langle x'y'\rangle\in{\cal
C}}{\rm sgn}(J_{x'y'})=\prod_{\langle x'y'\rangle\in{\cal C'}}{\rm
sgn}(J_{x'y'})$.

Just as in the ferromagnet, if percolation of a random cluster occurs in
the FK representation, the percolating cluster is unique (in each
realization of FK spins and bonds)~\cite{NS07,GKN92}.  In spite of this, RC
percolation alone is no longer sufficient to prove broken spin-flip
symmetry in the EA spin glass. This is because even in the presence of RC
percolation, it is unclear whether there exist any two sets of boundary
conditions that can alter the state of the spin at the origin from
arbitrarily far away.  Although the infinite cluster in any one RC
realization is unique, different RC realizations can have different paths
from $0\leftrightarrow\partial\Lambda_L$, and because of frustration this
can lead to different signs for $\sigma_0$. So percolation might still
allow for $\langle\sigma_0\rangle \to 0$ as $L\to\infty$, independently of
boundary condition.

Indeed, it is known that FK bonds percolate well above the spin glass
transition temperature.  For the three-dimensional Ising spin glass on the
cubic lattice, Fortuin-Kasteleyn bonds percolate at $\beta_{{\rm
FK},p}\approx 0.26$~\cite{ArCoPe91} while the inverse critical temperature
is believed to be $\beta_c=0.89\pm 0.03$~\cite{KaKoYo06}.  Near the spin
glass critical temperature, the giant FK cluster includes most of the sites
of the system.  For this reason, the Swendsen-Wang algorithm, though valid,
is inefficient for simulating spin glasses.

However, it is an open question as to whether the presence of ``single ''
(see below) FK percolation would lead to a slower, e.g. power-law, decay of
correlation functions even though the Gibbs state is unique.  If this were
to happen, then the onset of single FK percolation would imply a phase
transition in the spin glass, though not multiple Gibbs states and hence no
broken spin-flip symmetry.  This possibility was suggested in~\cite{NS07};
but to date, no evidence exists to support it.

Nonetheless, single FK percolation remains a {\it necessary\/} condition
for multiple (symmetry-broken) Gibbs phases in the spin glass, for the same
reason as for the ferromagnet.  A slightly stronger version of that
argument~\cite{N93} proves that the transition temperature for an EA spin
glass, if it exists, is bounded from above by the transition temperature in
the corresponding (disordered) ferromagnet.

The essential difference (from the point of view of FK percolation) between
the ferromagnet and the spin glass is the factor $1_U$ in~(\ref{eq:mufk}).
It is somewhat easier to discuss this factor in the context of finite
volumes.  Let $\Lambda_L$ denote the $L^d$ cube centered at the origin, and
let $\hat\Lambda_L$ denote the set of bonds $b=\langle x,y\rangle$ with
both $x$ and $y$ in $\Lambda_L$.  For models containing frustration, $U$ is
typically {\it not\/} all of $\{0,1\}^{\hat\Lambda_L}$, unlike for the
ferromagnet; it is the set of all {\it unfrustrated\/} bond configurations
(so that the product of couplings around any closed loop in such a
configuration is always positive).

So is there a way to extend the above considerations to arrive at a
sufficient condition for multiple Gibbs states in spin glass models?  We
begin by noting that FK clusters identify magnetization correlations; but
spin glass ordering is manifested by the Edwards-Anderson~(EA) order
parameter, and not the magnetization, becoming nonzero. The EA order
parameter $q_{EA}$ can be defined with respect to two independent replicas
of the system, each with the same couplings~$\{J_{xy}\}$.  Denoting the
spins in the two replicas by $\{\sigma_x\}$ and $\{\tau_x\}$, each taking
values $\pm 1$, $q_{EA}$ is defined in terms of the overlap,
\begin{equation}
Q \, = \, N^{\,-1}\sum_{\{x\}} \sigma_x \tau_x \ ,
\end{equation}
in the limit as the number of sites $N\to \infty$.  In general, $Q$ is a
random variable whose maximum possible value is $q_{EA}$, but in the case
where (in the limit $N\to \infty$) $\{\sigma_x\}$ and $\{\tau_x\}$ are
drawn from a single pure state, $Q$ takes on only the single value
$q_{EA}$.

Using this as a guide, it appears that one possibility, proposed
in~\cite{MNS07a,MNS07b,NS07}, for extending FK methods to spin glasses is
to use what might be called {\it double\/} FK percolation.  Here one
expands the sample space $\Omega$ to include two independent copies of the
bond occupation variables (for a given ${\cal J}$ configuration), and
defines the variable $r_{xy}=\omega_{xy}\omega'_{xy}$, where $\omega_{xy}$
and $\omega'_{xy}$ are taken from the two copies.  One then replaces
percolation of $\{\omega_{xy}\}$ in the single RC case with percolation of
$\{r_{xy}\}$.  It is not hard to see that this would be a sufficient
condition for the existence of multiple Gibbs phases (and consequently, for
a phase transition).
 
This is not the only way, however, to use ``double'' percolation of RC
clusters in some form to arrive at a condition for spin glass ordering; the
above describes what we denoted in the Introduction as the TRFK approach.
In the next subsection we present a closely related approach, the CRM
two-replica graphical representation, which requires a more lengthy
description.  Once this is done, we will discuss how both representations
can be used to illuminate the nature of phase transitions in short-range
spin glass models.

\subsection{The CMR Representation}
\label{subsec:CMR}

In describing these ``double FK'' representations, we will find it useful
to reformulate our description of the relevant measures slightly using a
joint spin-bond distribution introduced by Edwards and
Sokal~\cite{ES88}. The statistical weight ${\cal W}$ for the Edwards-Sokal
distribution on a finite lattice $\Lambda_L$ is
\begin{equation}
\label{eq:es}
{\cal W}(\{\sigma_x\},\{\omega_{xy}\};p)=p^{|\omega|}(1-p)^{N_b-|\omega|}
\Delta(\{\sigma_{x}\},\{\omega_{xy}\}) \, .
\end{equation}
Here $|\omega|=\sum_{\langle x,y\rangle} \omega_{xy}$ is the number of
occupied bonds and $N_b$ is the total number of bonds on the lattice.  The
factor $\Delta(\{\sigma_x\},\{\omega_{xy}\})$ is introduced to require that
every occupied bond is satisfied; it is defined by
\begin{equation}
\label{eq:delta}
\Delta(\{\sigma_x\},\{\omega_{xy}\})= \left\{ \begin{array}{l}
1 \mbox{   if  for every $xy$: }  
\omega_{xy}\sigma_x\sigma_y \geq 0
 \\0 \mbox{ otherwise} \, . \end{array} \right.
\end{equation}
With $p$ defined as in~(\ref{eq:pxy}) and with all $J_{xy}=J>0$, it is easy
to verify that the spin and bond marginals of the Edwards-Sokal
distribution are the ferromagnetic Ising model with coupling strength $J$
and the Fortuin-Kasteleyn random cluster model (cf.~(\ref{eq:mufk})),
respectively.

We can now adapt the above representation to the $\pm J$ Ising spin glass.
(With minor modifications, it can also be adapted to Gaussian and other
distributions for the couplings.)  The corresponding Edwards-Sokal weight
is the same as that given in Eq.\ (\ref{eq:es}).  The $\Delta$ factor must
still enforce the rule that all occupied bonds are satisfied,
\begin{equation}
\label{eq:essg}
\Delta(\{\sigma_x\},\{\omega_{xy}\};\{J_{xy}\})= \left\{ \begin{array}{l}
1 \mbox{   if  for every $xy$: }  
J_{xy}\omega_{xy}\sigma_x\sigma_y \geq 0
 \\0 \mbox{ otherwise} \, . \end{array} \right.
\end{equation}
so (cf.~(\ref{eq:mufk}))
$1_U=\max_{\{\sigma_x\}}\Delta(\{\sigma_x\},\{\omega_{xy}\};\{J_{xy}\})$. The
spin marginal of the corresponding Edwards-Sokal distribution is now the
Ising spin glass with couplings $\{J_{xy}\}$. But the relationship between
spin-spin correlations and bond connectivity is complicated by the presence
of antiferromagnetic bonds.  Now one has

\begin{eqnarray}
\langle \sigma_x \sigma_y \rangle=&\\
\mbox{Prob}\{x \mbox{ and} &y \mbox{ connected by even number of antiferromagnetic bonds}\}\nonumber\\
-\mbox{Prob}\{x \mbox{ and} &y \mbox{ connected by odd number of antiferromagnetic bonds}\}\nonumber.
\end{eqnarray}

The two-replica CMR graphical representation, introduced
in~\cite{CMR98,CRM98}, incorporates, in addition to the spin variables
$\{\sigma_x\}$ and $\{\tau_x\}$, two sets of bond variables
$\{\omega_{xy}\}$ and $\{\eta_{xy}\}$, each taking values in $\{0,1\}$.
Now the Edwards-Sokal weight is
\begin{multline}
\label{eqn:es}
{\cal W}(\{\sigma_x\},\{\tau_x\},\{\omega_{xy}\},\{\eta_{xy}\};\{J_{xy}\}) \\ = B_{\rm
\s}(\{\omega_{xy}\})B_{\rm \w}(\{\eta_{xy}\})
\Delta(\{\sigma_x\},\{\tau_x\},\{\omega_{xy}\};\{J_{xy}\})\Gamma(\{\sigma_x\},\{\tau_x\},\{\eta_{xy}\})
\end{multline}
where the $B$'s are Bernoulli factors for the two types of bonds,
\begin{eqnarray}
B_{\rm \s}(\{\omega_{xy}\})={\cal P}_{\rm \s}^{|\omega|} (1-{\cal P}_{\rm
\s})^{N_b-|\omega|} \\ B_{\rm \w}(\{\eta_{xy}\})={\cal P}_{\rm
\w}^{|\eta|}(1-{\cal P}_{\rm \w})^{N_b-|\eta|}
\end{eqnarray}
and the bond occupation probabilities are
\begin{eqnarray}
\label{eqn:blue}
{\cal P}_{\rm \s} & = & 1- \exp(-4\beta |J|) \\
\label{eqn:red}
{\cal P}_{\rm \w} & = & 1- \exp(-2\beta |J|).
\end{eqnarray}
So $B_{\rm \s}(\{\omega_{xy}\})=\mu_{\rm ind}(\{\omega_{xy}\})$. The
$\Delta$ and $\Gamma$ factors constrain where the two types of occupied
bonds are allowed,
\begin{eqnarray}
\lefteqn{
\Delta(\{\sigma_x\},\{\tau_x\},\{\omega_{xy}\};\{J_{xy}\})
}\nonumber \\ 
&=&\left\{ \begin{array}{l} 1 \mbox{ if for every $xy$: }
J_{xy}\omega_{xy}\sigma_x \sigma_y \geq 0 \mbox{ and }
J_{xy}\omega_{xy}\tau_x \tau_y\geq 0
\\0 \mbox{ otherwise} \, , \end{array} \right.\\
\lefteqn{
\Gamma(\{\sigma_x\},\{\tau_x\},\{\eta_{xy}\})
}\nonumber \\ 
&=&
\left\{ \begin{array}{l} 1
\mbox{ if for every $xy$: } 
\eta_{xy}\sigma_x \sigma_y\tau_x\tau_y \leq 0
\\0 \mbox{ otherwise} \, . \end{array} \right.
\end{eqnarray}

We refer to the $\omega$-occupied bonds as ``blue'' and the $\eta$-occupied
bonds as ``red''.  The $\Delta$ constraint says that blue bonds are allowed
only if the bond is satisfied in both replicas.  The $\Gamma$ constraint
says that red bonds are allowed only if the bond is satisfied in exactly
one replica.

It is straightforward to verify that the spin marginal of the two-replica
Edwards-Sokal weight is that for two independent Ising spin glasses with
the same couplings,
\begin{equation}
\label{eqn:spinmarg}
\sum_{\{\omega\}\{\eta\}} {\cal
W}(\{\sigma_x\},\{\tau_x\},\{\omega_{xy}\},\{\eta_{xy}\};\{J_{xy}\})=
\mbox{const} \times \exp \left[ \beta \sum_{\langle xy\rangle} J_{xy}
(\sigma_x\sigma_y+\tau_x\tau_y)\right]
\end{equation}

Connectivity by occupied bonds in the two-replica representation is related
to correlations of the local spin glass order parameter,
\label{sec:pgr}
\begin{equation}
Q_x = \sigma_x\tau_x .
\end{equation}
It is straightforward to verify that 
\begin{eqnarray}
\label{eq:trcor}
\langle Q_x Q_y \rangle = 
\mbox{Prob}\{x \mbox{ and } y \mbox{ connected by even number of red bonds}\}\\
-\mbox{Prob}\{x \mbox{ and } y \mbox{ connected by odd number of red bonds}\}.\nonumber
\end{eqnarray}
As in the case of the FK representation, a minus sign complicates the
relationship between correlations and connectivity but in a conceptually
different way.  The second term in Eq.\ (\ref{eq:trcor}) is independent of
the underlying couplings in the model and is present for both spin glasses
and ferromagnetic models.

As noted earlier, for ferromagnets (in the absence of
non-translation-invariant boundary conditions), the signature of ordering
is a {\it single\/} percolating FK cluster.  For spin glasses, the
situation is more complicated
as there can be more than one percolating cluster. 
However, if the CMR graphical representation displays a
single percolating blue cluster of largest density, one can similarly show
broken symmetry, for either EA or SK models. This is because one can impose
``agree'' or ``disagree'' boundary conditions between those $\sigma_x$ and
$\tau_x$ boundary spins belonging to the maximum density blue network.  In
the infinite volume limit, these two boundary conditions give different
Gibbs states for the $\sigma$-spin system (for fixed $\tau$) related to
each other by a global spin flip (of $\sigma$).  

It is a separate matter, however, to relate in a simple way the EA order
parameter to the density difference in blue (or more generally, doubly
percolating) clusters.  Intuitively, it seems that there should be a simple
correspondence, and in fact such a relation is easy to show for the SK
model: here the overlap $Q$ is exactly the difference in density between
two percolating blue clusters~\cite{MNS07a}.  But it is not immediately
obvious that a similar density difference can be simply related to the EA
order parameter (although the arguments above imply that here also a
density difference implies a nonzero $Q$.)  This is because in a
two-replica situation, it is not immediately obvious that there is no
contribution from finite clusters in short-range models.

For the TRFK representation, similar reasoning shows that the occurrence of
exactly two doubly-occupied percolating FK clusters with different
densities implies broken symmetry for the spin system~\cite{NS07} and that
$Q$ should equal (and once again, does equal in the SK model) the density
difference.  In~\cite{MNS07a} we presented preliminary numerical evidence
that there is such a nonzero density difference below the spin glass
transition temperature for the $d=3$ EA $\pm J$ spin glass (for both the
TRFK and CMR representations).  We will discuss these results further in
Sec.~\ref{sec:numerical}.  But now we turn to a review of some rigorous
results, particularly for the SK model, that appeared in~\cite{MNS07a} and
\cite{MNS07b}.

\section{Rigorous Results}
\label{sec:rigorous}

The SK model permits a fairly extensive rigorous analysis of both the CMR
and TRFK representations which, when combined with other known results,
permits a sharp picture to emerge of the connection between double FK
percolation (we use this more general term to refer to any representation
that relies on two FK replicas, such as CMR or TRFK), and a phase
transition to a low-temperature spin glass phase with broken spin-flip
symmetry.  The surprising feature to emerge from this analysis is that in
the CMR representation, there already exist well {\it above\/} the
transition temperature two percolating networks of blue bonds, of equal
density and in all respects macroscopically indistinguishable.  Below the
critical temperature $T_c$, the indistinguishability is lifted: the two
infinite clusters assume different densities.

On the other hand, there is no double FK percolation at all above $T_c$ in
the TRFK representation, but below $T_c$ there are again two percolating
double clusters of unequal density.  

We do not yet know whether this difference in the two pictures above $T_c$
persists in the EA model, but numerical evidence (to be discussed in the
next section) so far appears to indicate that it does not: in the EA spin
glass, both representations seem to behave similarly to CMR in the SK
model.

The SK Hamiltonian for an $N$-spin system is 
\begin{equation}
\label{eq:SK}
{\cal H}_N=-{\frac{1}{\sqrt{N}}} 
\sum_{1\le i<j\le N}J_{ij}\sigma_i\sigma_j\ ,
\end{equation}
where $i,j=1,\ldots N$ are vertices on a complete graph.  The couplings
$J_{ij}$ are i.i.d. random variables chosen from a probability measure
$\rho$ satisfying the following properties:

\noindent 1) The distribution is symmetric: $\rho(u)=\rho(-u)$.

\noindent 2) $\rho$ has no $\delta$-function at $u=0$.

\noindent 3) The moment generating function is finite:
$\int_{-\infty}^\infty d\rho(u)\ e^{-t|u|}<\infty$ for all real $t$.

\noindent 4) The second moment of $\rho$ is finite; we normalize it to one:
$\int_{-\infty}^\infty d\rho(u)\ u^2=1$.

So the results of this section hold for Gaussian and many other
distributions (although not diluted ones, because of requirement (2)), but
the analysis is simplest for the $\pm J$ model, to which we confine our
attention for the remainder of this section. We therefore assume that
$J_{ij}=\pm 1$ with equal probability; for this distribution (or any other
satisfying property (3)) $\beta_c=1$~\cite{BY86,SK75}.

It is relatively easy to see, using a heuristic argument presented
in~\cite{MNS07a}, at what temperature a single FK cluster will form.  The
SK energy per spin, $u$, is given above the critical temperature by
$u=-\beta/2$.  Therefore, for large $N$ the fraction $f_s$ of satisfied
edges is
\begin{equation}
\label{eq:fracsat}
f_s \sim \frac{1}{2} - u N^{-1/2} \ .
\end{equation} 

From~(\ref{eq:pxy}) it follows that a fraction ${\cal P}_{\rm FK}= 1-
\exp(-2 \beta N^{-1/2})\approx 2 \beta N^{-1/2}$ of satisfied edges are
occupied.  According to the theory of random graphs
(see~\cite{Bollobas01}), a giant cluster forms in a random graph of $N$
vertices when a fraction $x/N$ of edges is occupied with $x>1$, and there
is then a single giant cluster~\cite{ER60}.  So if edges are satisfied
independently (which of course they're not --- this is why this argument is
heuristic only), then single replica FK giant clusters should form when
$\beta = x N^{-1/2}$ when $x>1$.  The single replica FK percolation
threshold is therefore at
\begin{equation}
\label{eq:fkperc}
\beta_{{\rm FK},p}=N^{-1/2} \, ,
\end{equation}
and above this threshold, there should be a single giant FK cluster.

This simple argument can be made rigorous~\cite{NS07}, but the basic ideas
are already displayed above.  The rigorous argument obtains upper and lower
bounds for the conditional probability that an edge $\{x_0 y_0\}$ is
satisfied, given the satisfaction status of all the other edges. If these
bounds are close to each other (for large $N$) then treating the satisfied
edges as though chosen independently can be justified.  We now describe
that argument.

To avoid the problem of non-independence of satisfied edges, we fix all
couplings $J_{xy}$ but one, which we denote $J_{x_0 y_0}$, and ask for the
conditional probability of its sign given the configuration of all other
couplings and all spins $\sigma_x$. The ratio $Z_{+}/Z_{-}$ of the
partition functions with $J_{x_0 y_0}=+N^{-1/2}$ and $J_{x_0
y_0}=-N^{-1/2}$ satisfies
\begin{equation}
\exp(-2 \beta N^{-1/2}) \, \leq \, |Z_{+}/Z_{-}| \, \leq
\exp(2 \beta N^{-1/2}) \, .
\end{equation}
The conditional probabilities $P_{\pm}$ that $J_{x_0 y_0}=\pm N^{-1/2}$
therefore satisfy
\begin{equation}
e^{-4\beta/\sqrt{N}} \leq e^{-2\beta/\sqrt{N}}|Z_{-}/Z_{+}|
\leq P_{+} / P_{-}\leq e^{2\beta/\sqrt{N}}|Z_{-}/Z_{+}|
\leq e^{4\beta/\sqrt{N}} \, .
\end{equation}

Let $P_s$ ($P_u$) be the conditional probability for any edge $\{x_0 y_0\}$
to be satisfied (unsatisfied) given the satisfaction status of all other
edges.  These must then be bounded as follows:
\begin{equation}
e^{-4\beta/\sqrt{N}} \leq P_s/P_u \leq e^{4\beta/\sqrt{N}} \, ,
\end{equation}
and therefore

\begin{equation}
\frac{1}{2}-{\mathrm O}(\beta/\sqrt{N}) = (e^{4\beta/\sqrt{N}}+1)^{-1}
\leq P_s \leq (e^{-4\beta/\sqrt{N}}+1)^{-1} =\frac{1}{2}+{\mathrm O} (\beta/\sqrt{N})\,.
\end{equation}
One now obtains rigorously the same conclusions as before --- i.e.,
(\ref{eq:fkperc}) is valid with a single giant FK cluster
for $\beta = \beta_N \geq x N^{-1/2}$ with any $x>1$.

Essentially the same argument is used for the more interesting case of
double percolation.  Here there are two spin replicas denoted by $\sigma$
and $\tau$. We are now interested in percolation of doubly satisfied edges;
the spins at the vertices of each such edge must satisfy $\sigma_x \tau_x
= \sigma_y \tau_y$ (and then will be satisifed for exactly one of the two
signs of $J_{xy}$).  We note an immediate difference between the CMR and
TRFK representations: for the former, doubly satisfied edges are occupied
with probability ${\cal P}_{\rm \tr} = 1- \exp(-4 \beta N^{-1/2}) \sim 4
\beta N^{-1/2}$, while for the latter, ${\cal P}_{\rm \dfk} = [1- \exp(-2
\beta N^{-1/2})]^2 \sim 4\beta^2/N$.  It seems likely that this difference
occurs only for the SK model; the various factors of $N$ are absent in the
EA model.

We can proceed much as in the single-replica case by dividing all
$(\sigma,\tau)$ configurations into two sectors --- the {\it agree\/}
(where $\sigma_x = \tau_x$) and the disagree sectors (where $\sigma_x =
-\tau_x$). We also denote by $N_a$ and $N_d$ the numbers of sites in the
sectors and denote by $D_a = N_a/N$ and $D_d= N_d/N$ the sector densities
(so that $D_a+D_d=1$). The spin overlap $Q$ is then just
\begin{equation}
Q = \frac{1}{N} \sum_x \sigma_x \tau_x = \frac{N_a-N_d}{N} = D_a -D_d\, .
\end{equation}
For $\beta \leq \beta_c =1 $, $Q\to 0$ as $N \to \infty$ (because the EA
order parameter is zero in the paramagnetic phase) while for $\beta >
\beta_c =1 $, $\overline{ \langle Q^2 \rangle} >0$ as $N \to \infty$, where
$\overline{(\cdot)}$ denotes an average over couplings.  So it must be that
$D_a=D_d$ for $\beta\leq 1$ while $D_a\ne D_d$ for $\beta>1$.

The arguments for the single replica case can be repeated separately within
the agree and disagree sectors. Letting $\bar P_\pm$ denote the conditional
probabilities that $J_{x_0 y_0} = \pm N^{-1/2}$ given the other $J_{x y}$'s
and all $\sigma_x$'s and $\tau_x$'s, we have within either of the two
sectors that
\begin{equation}
e^{-8\beta/\sqrt{N}} \leq e^{-4\beta/\sqrt{N}}|Z_{-}/Z_{+}|^2
\leq {\bar P_{+}} / {\bar P_{-}}\leq e^{4\beta/\sqrt{N}}|Z_{-}/Z_{+}|^2
\leq e^{8\beta/\sqrt{N}}
\end{equation}
so that the conditional probability {\it within a single sector\/} $P_{ds}$
for ${x_0 y_0}$ to be doubly satisfied is $(1/2) + {\mathrm O}(\beta
N^{-1/2})$. For $\beta \leq \beta_c$, we have $D_a =1/2$, $ D_d =1/2$ (in
the limit $N \to \infty$) and so in either sector, double FK percolation is
approximately a random graph model with $ N/2$ sites and bond occupation
probability $(1/2) 4 \beta^2 N^{-1} = \beta^2 (N/2)^{-1}$; thus double FK
giant clusters do not occur for $\beta^2 \leq 1$ in the TRFK
representation.

In the CMR representation, blue percolation corresponds to bond
occupation probability $(1/2) 4 \beta N^{-1/2}$ $= \beta N^{1/2}(N/2)^{-1}$
and so the threshold for blue percolation is given by $\beta_{{\rm
CMR},p}=N^{-1/2}$.  But now there are {\it two\/} giant clusters, one in
each of the two sectors, and as noted above they satisfy $D_a=D_d$ when
$\beta \leq \beta_c =1$. For $\beta > \beta_c$, $D_a \neq D_d$. Since
$\beta N^{1/2} \to \infty$ for $\beta > \beta_c$ (indeed for any fixed
$\beta>0$), it follows from random graph theory that each giant cluster
occupies the entire sector so that $D_a$ and $D_d$ are also the cluster
percolation densities of the two giant clusters.

In the case of two-replica FK percolation for $\beta > \beta_c$, let us
denote by $D_{max}$ and $D_{min}$ the larger and smaller of $D_a$ and
$D_d$, so that $D_{max}+D_{min}=1$ and $D_{max}-D_{min}=Q$. Then for $\beta
> \beta_c$, the bond occupation probability in the larger sector is
$\beta^2(N/2)^{-1} = 2 \beta^2 D_{max} (D_{max}N)^{-1}$ with $2 \beta^2
D_{max}>1$ and there is a (single) giant cluster in that larger
sector. There will be another giant cluster (of lower density) in the
smaller sector providing $2 \beta^2 D_{min}=\beta^2 (1-Q) >1$. Since $Q\leq
q_{EA}$, for this to be the case it suffices if for $\beta > \beta_c$,
\begin{equation}
q_{EA} < 1 - \frac{1}{\beta^2} \ .
\end{equation}
The estimated behavior of $q_{EA}$ both as $\beta \to 1+$ and as $\beta \to
\infty$~\cite{BY86} suggests that this is always valid. In any case, we
have rigorously proved that there is a unique maximal density double FK
cluster for $\beta > \beta_c$.

An important feature of spin glass order is ultrametricity, which is
believed to be true at least for the SK model~\cite{MPSTV84}, although
it has not yet been proved rigorously. For the spin overlaps coming from
three replicas put into rank order, $Q_{(1)} \geq Q_{(2)} \geq Q_{(3)}$,
ultrametricity is the property that $Q_{(2)} = Q_{(3)}$. The issue of
percolation signatures for ultrametricity in the SK model is treated at
length in~\cite{MNS07b}. We only mention here one of those signatures,
which concerns the {\it four\/} percolating clusters that arise in
a CMR representation of three replicas when considering bonds that are 
simultaneously blue {\it both\/} for replicas one and two as well as
for replicas one and three. Denoting the four cluster densities in rank order
as $x_{(1)} \geq x_{(2)} \geq x_{(3)} \geq x_{(4)}$, the percolation
version of the ultrametric
property is that $x_{(1)}>x_{(2)}$ {\it and\/} $x_{(3)} = x_{(4)}$.
The case $x_{(2)} =x_{(3)}$ (resp., $x_{(2)} > x_{(3)}$) corresponds
to $Q_{(1)} > Q_{(2)}$ (resp, $Q_{(1)} = Q_{(2)}$).
See~~\cite{MNS07b} for more details.

\section{Numerical Methods and Results}
\label{sec:numerical}
We have carried out numerical simulations to explore the properties of two-replica FK representations described in the previous sections for the case of the three-dimensional $\pm J$ EA spin glass.  The Monte Carlo algorithm that we use also makes use of the CMR representation in addition to parallel tempering and Metropolis sweeps.  Similar methods have been previously applied by  Swendsen and Wang~\cite{SwWa86,WaSw88,WaSw05} and others~\cite{Houdayer01,jorg05,jorg06}.  The algorithm is described in more detail in \cite{MNS07a}, here we provide some additional details about the CMR component of the algorithm.  The CMR cluster algorithm alternates between ``bond moves'' and ``spin moves.'' The bond move takes a pair of  EA spin configurations $\{\sigma_x\}$ and $\{\tau_x\}$, both in the same realization of disorder, $\{J_{xy}\}$, and populates the bonds of the lattice with red and blue CMR bonds.  Each bond that is satisfied in both replicas is occupied with a blue bond with probability ${\cal P}_{\rm \s}$ and each bond that is satisfied in only one replica is occupied with a red bond with probability ${\cal P}_{\rm \w}$, defined in (\ref{eqn:blue}) and (\ref{eqn:red}), respectively.   
The bond move is described by the conditional probability, $\T(\{\omega_{xy}\},\{\eta_{xy}\} | \{\sigma_x\},\{\tau_x\};\{J_{xy}\})$ for bond configuration $\{\omega_{xy}\},\{\eta_{xy}\}$ given spin configuration $\{\sigma_x\},\{\tau_x\}$.  To simplify the notation, we omit the lattice indices and the dependence on $J_{xy}$ in the following equations except where explicitly needed. For the bond move we have,
\begin{equation}
\label{eqn:tomsig}
\T(\{\omega\},\{\eta\} | \{\sigma\},\{\tau\})=
\frac{
B_{\rm \s}(\{\omega\})B_{\rm \w}(\{\eta\})
\Delta(\{\sigma\},\{\tau\},\{\omega\})
\Gamma(\{\sigma\},\{\tau\},\{\eta\})
}{
\sum_{\{\omega'\}\{\eta'\}} 
B_{\rm \s}(\{\omega'\})B_{\rm \w}(\{\eta'\})
\Delta(\{\sigma\},\{\tau\},\{\omega'\})
\Gamma(\{\sigma\},\{\tau\},\{\eta'\})
 }.
\end{equation}
The spin move takes a blue and red CMR bond configuration and produces a pair of spin configurations $\{\sigma_x\}$ and $\{\tau_x\}$.  All spin configurations obeying the constraints that blue bonds are doubly satisfied and red bond are singly satisfied have the same probability.  The spin move is described by the conditional probability, $\T( \{\sigma_x\},\{\tau_x\} |\{\omega_{xy}\},\{\eta_{xy}\})$:
\begin{equation}
\T( \{\sigma\},\{\tau\} |\{\omega\},\{\eta\})=
\frac{
\Delta(\{\sigma\},\{\tau\},\{\omega\})
\Gamma(\{\sigma\},\{\tau\},\{\eta\})
}{
\sum_{\{\sigma'\},\{\tau'\}} 
\Delta(\{\sigma'\},\{\tau'\},\{\omega\})
\Gamma(\{\sigma'\},\{\tau'\},\{\eta\})
}.
\end{equation}

A bond move followed by a spin move constitututes a single sweep of the CMR cluster  algorithm.  We now demonstrate the validity of the CMR algorithm by showing that detailed balance and ergodicity are satisfied.  Ergodicity is clearly satisfied since there is a non-vanishing probability that no red or blue bonds will be created during a bond move so that the spin move can then produce any spin configuration.  Detailed balance with respect to the equilibrium spin distribution can be stated as 
\begin{multline}
\label{eqn:detailed}
\sum_{\{\omega''\},\{\eta''\}}\T( \{\sigma\},\{\tau\} |\{\omega''\},\{\eta''\})
\T(\{\omega''\},\{\eta''\} | \{\sigma'\},\{\tau'\}) \, \times
\exp \left[ \beta \sum_{\{xy\}} J_{xy}
(\sigma'_x\sigma'_y+\tau'_x\tau'_y)\right]=\\
\sum_{\{\omega''\},\{\eta''\}}\T( \{\sigma'\},\{\tau'\} |\{\omega''\},\{\eta''\})
\T(\{\omega''\},\{\eta''\} | \{\sigma\},\{\tau\}) \, \times
\exp \left[ \beta \sum_{\{xy\}} J_{xy}
(\sigma_x\sigma_y+\tau_x\tau_y)\right] .
\end{multline}
This equation must hold for all choices of $\{\sigma_x\}$, $\{\tau_x\}$, $\{\sigma'_x\}$ and $\{\tau'_x\}$. Note that by (\ref{eqn:es}) and (\ref{eqn:spinmarg}) the denominator in (\ref{eqn:tomsig}) is simply the Boltzmann weight for two independent EA spin glasses, cancelling the same factors in the numerator of the detailed balance equation. Thus, the LHS of (\ref{eqn:detailed}) can be written as
\begin{multline}
\sum_{\{\omega''\},\{\eta''\}}
\frac{
\Delta(\{\sigma\},\{\tau\},\{\omega''\})
\Gamma(\{\sigma\},\{\tau\},\{\eta''\})
}{
\sum_{\{\sigma''\},\{\tau''\}} 
\Delta(\{\sigma''\},\{\tau''\},\{\omega''\})
\Gamma(\{\sigma''\},\{\tau''\},\{\eta''\})
} \times \\
B_{\rm \s}(\{\omega\})B_{\rm \w}(\{\eta\})
\Delta(\{\sigma'\},\{\tau'\},\{\omega''\})
\Gamma(\{\sigma'\},\{\tau'\},\{\eta''\})
\end{multline}
Note that this expression is symmetric under the exchange of $\{\sigma_x\}$, $\{\tau_x\}$ and $\{\sigma'_x\}$, $\{\tau'_x\}$ and thus equal to the RHS of (\ref{eqn:detailed}), demonstrating that detailed balance holds.  The bond configurations observed after the bond move are, in the same fashion, easily shown to be equilibrium CMR bond configurations as described by the bond marginal of the Edwards-Sokal weight (\ref{eqn:es}).  

Although the CMR algorithm correctly samples equilibrium configurations of EA spin glasses, in 3D it is not efficient.  To obtain a more efficient algorithm, we also make use of parallel tempering and Metropolis sweeps.  The CMR algorithm employs two replicas at a single temperature while parallel tempering exchanges replicas at different temperatures.  Here we use 20 inverse temperatures equally spaced between $\beta=0.16$ to $\beta=0.92$.
 The phase transition temperature of the system was recently measured as
$\beta_c=0.89\pm 0.03$ \cite{KaKoYo06}.
A single sweep of the full algorithm consists of a CMR cluster sweep for each pair of replicas at each temperature, a parallel tempering exchange between replicas at each pair of neighboring temperatures and a Metropolis sweep for every replica.

We simulated the three-dimensional $\pm J$ Edwards-Anderson model on skew
periodic cubic lattices for system sizes $6^3$, $8^3$, $10^3$ and $12^3$.
For each size we simulated 100 realizations of disorder for 50,000 Monte Carlo sweeps of which the first
$1/4$ of the sweeps were for equilibration and the remaining $3/4$ for data
collection.   The quantities that we
measure are the fraction of sites in the largest \s cluster, ${\cal C}_1$
and second largest \s cluster, ${\cal C}_2$ and the number of \s \tr
wrapping cluster, $w_{\rm \tr}$, and the number of \dfk ``wrapping''
clusters, $w_{\rm \dfk}$.  A cluster is said to wrap if it is connected
around the system in any of the three directions.

\begin{figure}
 \includegraphics{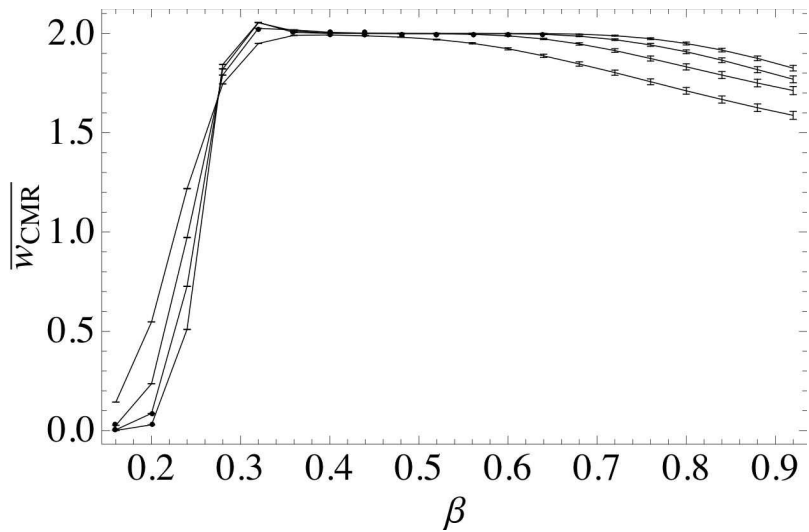}
\caption{Average number of wrapping \tr clusters, $\overline{w_{\rm CMR}}$ vs.\  $\beta$ for the 3D EA model.}
\label{fig:wrap}       \end{figure}

Figure \ref{fig:wrap} shows the average number $\overline{w_{\rm CMR}}$ of
\tr blue wrapping clusters 
as a function of inverse
temperature $\beta$.  The curves are ordered by system size with largest
size on the bottom for the small $\beta$ and on top for large $\beta$.  The
data suggests that there is a percolation transition at some $\beta_{{\rm
CMR},p}$.  For $\beta> \beta_{{\rm CMR},p}$ there are {\em two} wrapping
clusters while for $\beta<\beta_{{\rm CMR},p}$ there are none. Near and
above the spin glass transition at $\beta_c \approx 0.89$ the expected
number of wrapping clusters falls off but the fall-off diminishes as system
size increases.  This figure suggests that in the large size limit there
are exactly two spanning clusters near the spin glass transition both above
and below the transition temperature.  

\begin{figure}
 \includegraphics{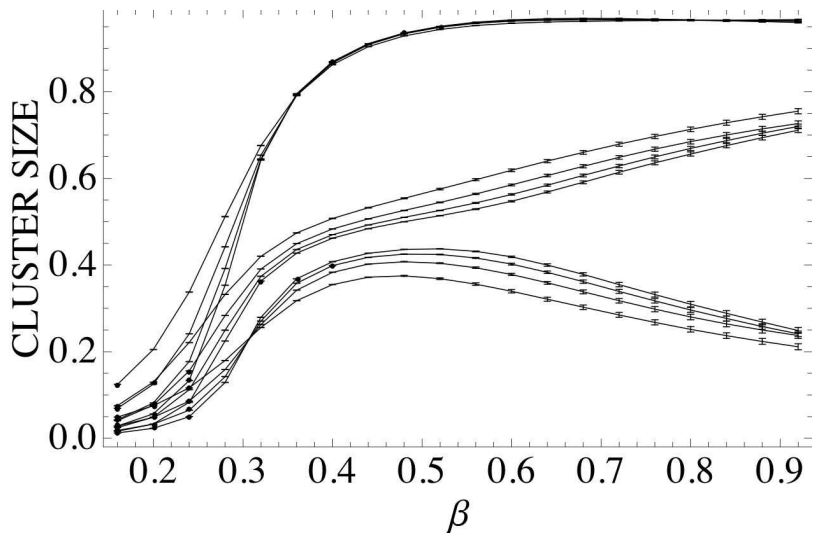}
\caption{${\cal C}_1$ (middle set), ${\cal C}_2$ (bottom set) and ${\cal C}_1+{\cal C}_2$ (top set) vs.\
$\beta$ for the  \tr graphical representation for the 3D EA model.}
\label{fig:clustersize}       
\end{figure}

Figure \ref{fig:clustersize} shows the fraction of sites in the largest \tr
blue cluster, ${\cal C}_1$, second largest \tr blue cluster, ${\cal C}_2$
and the sum of the two, ${\cal C}_1+{\cal C}_2$. The middle set of four
curves is ${\cal C}_1$ for sizes $6^3$, $8^3$, $10^3$ and $12^3$, ordered
from top to the bottom at $\beta=0.5$.  The bottom set of curves is ${\cal
C}_2$ with systems sizes ordered from smallest on bottom to largest on top
at $\beta=0.5$.  The difference between the fraction of sites in the two
largest clusters, ${\cal C}_1-{\cal C}_2$ is approximately the spin glass
order parameter.  As the system size increases, this difference decreases
below the transition suggesting that ${\cal C}_1={\cal C}_2$ for
$\beta<\beta_c$ in the thermodynamic limit.  On the other hand, the sum of
the two largest clusters is quite constant independent of system size. Near
the transition, approximately 96\% of the sites are in the two largest
clusters.

The large fraction of sites in the two largest clusters makes the \tr
cluster moves inefficient.  If all sites were in the two largest clusters
then the cluster moves would serve only to flip all spins in one or both
clusters or exchange the identity of the two replicas.  Equilibration
depends on the small fraction of spins that are not part of the two largest
clusters. 

\begin{figure}
\includegraphics{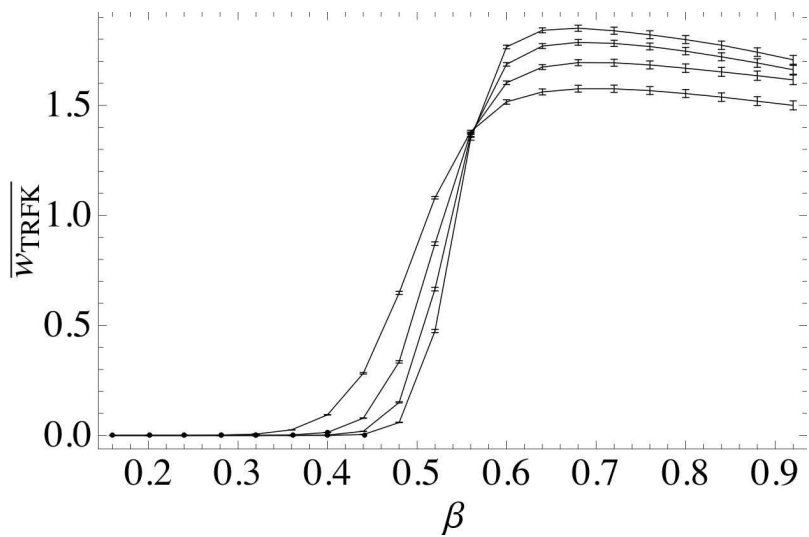}
\caption{Average number of doubly occupied wrapping Fortuin-Kasteleyn
clusters, $\overline{w_{\rm \dfk}}$ vs.\ $\beta$ for the 3D EA model.}
\label{fig:fkwrap}       
\end{figure}

Figure \ref{fig:fkwrap} show the average number of wrapping \dfk clusters 
$\overline{w_{\rm \dfk}}$
as a function of inverse temperature.  The largest system size is on the
bottom for the small $\beta$ and on top for the large $\beta$.  As for the
case of \tr clusters, the data suggests a transition at some $\beta_{{\rm
\dfk},p}$ from zero to two wrapping \dfk clusters.  Although the number of
\dfk wrapping clusters is significantly less than two for all $\beta$ and
all system sizes, the trend in system size suggests that it might approach
two for large systems and $\beta>\beta_{{\rm \dfk},p}$. 

The percolation signature for both \tr and \dfk clusters is qualitatively
similar in three dimensions.  In both cases two giant clusters with
opposite values of the local order parameter appear at a temperature
substantially above the phase transition temperature.  In the high
temperature phase, the two giant clusters have the same density and the
phase transition is marked by the onset of different densities of the two
clusters. The numerical evidence, however, suggests that the transition from equal to unequal giant cluster densities is quite broad for the small system sizes explored here.

\section{Discussion}

We begin by noting that our results, even for the SK infinite-range spin
glass, are not used to prove a phase transition (which has already been
proved using other techniques~\cite{ALR87}).  Rather, we use as an input
the fact that such a transition exists and that the EA order parameter is
zero above $T_c$ and nonzero below, and then show that the transition
coincides exactly with the onset of a density difference in doubly occupied
FK clusters.  Numerical evidence indicates that something similar is
occurring in the EA model.  Together these complementary approaches shed
light on the nature of the spin glass transition, especially from a
geometric viewpoint, and suggest a possible framework through which a spin
glass phase transition in realistic spin glass models can be finally
proved.

We now summarize our results.  We have introduced a random cluster approach
for studying phase transitions and broken symmetry in spin glasses, both
short- and infinite-range.  We have shown that, unlike for ferromagnetic
models, single FK percolation is a necessary but not sufficient condition
for broken spin flip symmetry.  However, double FK percolation (with a
unique largest cluster) is sufficient and probably necessary.  (More
precisely, it {\it is\/} necessary in the SK model, because otherwise the
EA order parameter is zero; and it is probably necessary in the EA model.)

In the SK model, there is a difference above $T_c$ between the CMR and TRFK
approaches, but not below.  In the former, there already is double
percolation below $\beta_c$, above an onset inverse temperature of
$\beta_{{\rm CMR}, p}= N^{-1/2}$, below which there are no giant clusters.
Between this inverse temperature and $\beta_c$ (which equals one in the
class of models we study) there are exactly two giant clusters of equal
density, which become $1/2$ in the $N\to\infty$ limit. 

More importantly, there is a second transition occurring at exactly the SK
spin glass critical value $\beta_c =1$. Above this threshold, the two giant
clusters take on unequal densities, whose sum is one (i.e., every bond and
spin belong to one of the two giant clusters).  It could be the case at
some even higher $\beta$ there is only a single giant cluster, but our
methods so far are unable to determine whether this is the case.

For TRFK percolation there are no giant clusters above $\beta_{{\rm
TRFK},p} = 1$.  For $\beta > 1$ there are exactly two giant clusters with
unequal densities, and the picture then becomes similar to that of the CMR
representation.

The numerical simulations of the 3D EA model suggest a scenario similar
to what we find in the SK model.  We observe a sharp percolation transition
for both the CMR and TRFK representations at a temperature well above 
the spin glass transition temperature.  At this percolation transition,
two giant clusters form with opposite values of the EA order parameter.
These clusters are nearly equal in density  and together occupy most of
the system.  As the temperature is decreased toward the presumed location
of the spin glass transition, the density of the two clusters becomes 
increasingly unequal.  Although this transition in the density of the 
two largest clusters appears quite rounded for the small systems 
investigated numerically, we believe it is sharp in the thermodynamic 
limit.  The results for both the SK model and the connection between the
EA order parameter and the density difference between giant clusters 
strongly suggests that the spin glass transition in finite dimensions 
is marked by the onset of this density difference in both the CMR 
and TRFK representations.  These results provide an interesting geometric
avenue for investigating the phase transition in the EA model and related
models with frustration such as the random bond Ising model or the Potts
spin glass.

\begin{acknowledgements} 
JM was supported in part by NSF DMR-0242402. CMN was supported in part by
NSF DMS-0102587 and DMS-0604869.  DLS was supported in part by NSF
DMS-0102541 and DMS-0604869.  Simulations were performed on the Courant
Institute of Mathematical Sciences computer cluster. CMN and DLS thank the
organizers of the 
2007 Paris Summer School ``Spin Glasses'', from which these lectures
were adapted. JM and DLS thank the Aspen Center for Physics, where some of
this work was done.
\end{acknowledgements}



\small\def\em{\it} \newcommand{\noopsort}[1]{} \newcommand{\printfirst}[2]{#1}
  \newcommand{\singleletter}[1]{#1} \newcommand{\switchargs}[2]{#2#1}

\end{document}